\documentclass[12pt]{article} 
\usepackage{amssymb}
\usepackage{graphicx}
\begin{document} 
\begin{center}
      {\large \bf  Ultraperipheral vs ordinary nuclear interactions  } 

\vspace{0.5cm}                   
{\bf I.M. Dremin}

\vspace{0.5cm}              
        Lebedev Physical Institute, Moscow 119991, Russia

\end{center}


\begin{abstract}
It is argued that the cross sections of ultraperipheral interactions of heavy
nuclei can become comparable in value to those of their ordinary hadronic 
interactions at high energies. Simple estimates of corresponding
``preasymptotic energy thresholds'' are provided. The~method of equivalent photons
is compared with the perturbative approach. The~situation at NICA/FAIR 
energies is discussed.
\end{abstract}

\section{Introduction}
The cross sections of both ultraperipheral and ordinary hadronic nuclear
interactions increase with the rise in collision energies. The~rate of the 
increase is higher for ultraperipheral collisions with large impact parameters
where the electromagnetic fields of colliding charged objects play a 
dominant role. Surely, at lower energies, their strength is weaker
compared to effects due to strong hadronic (quark-gluon) interactions.
Thus, starting from smaller initial values, ultraperipheral cross sections
have the chance to overcome at higher energies the contribution of ordinary 
processes if the electromagnetic fields between colliding heavy nuclei are 
strong enough.

Landau and Lifshitz were the first to show \cite{lali} that the cross section
$\sigma $ for the production of an electron and positron in ultraperipheral
nuclear ($A$) collisions increases with the cube of the logarithm of the 
energy $E$:
\begin{equation}
\sigma (AA\rightarrow AAe^-e^+) \propto \ln ^3 \gamma ,
\label{gamma}
\end{equation}
where $\gamma = E/m $ is the Lorentz boost\footnote{The asymptotic dependence 
is not changed if the laboratory frame, used in Ref. \cite{lali}, is replaced by 
the center of mass system.}. It is important that the prefactor in
Equation (\ref{gamma}) is proportional to $Z^4$ where $Z$ is the nucleus charge.
Therefore, the~cross sections of ultraperipheral collisions of heavy nuclei
are strongly enhanced. This result was obtained by considering the Dirac
equation and new ideas of the creation of positrons from the Dirac sea.

In these collisions, the~two colliding protons or nuclei interact 
electromagnetically but not hadronically. They effectively miss each other
interacting by their photon clouds only, which create the electron-positron
pairs. No nuclear transitions appear at small transferred momenta. 

Such interactions were first considered by Fermi \cite{fermi} almost a century 
ago. Ten years later, the~method of equivalent photons \cite{lali, weiz, will}
was developed and effectively used for quantitative estimates. The~photons
in the clouds of fast moving nuclei can be considered almost real because their
energy is much higher than their virtuality (the four-momenta squared). 
Unfortunately, this approximation is limited by asymptotic formulas, such as 
Equation (\ref{gamma}). To get the preasymptotic behavior one should calculate the 
factor $\gamma _0$ in the ratio $\gamma /\gamma _0$ within the logarithm
in Equation (\ref{gamma}). The~condition $\gamma \gg \gamma _0$ is necessary
for asymptotics. To perform the program one has to use some knowledge of 
the structure of the colliding objects, the~masses of produced particles etc. 
The new parameters enter the game.

\begin{figure}
\centering
\includegraphics[width=13cm, height=9cm]{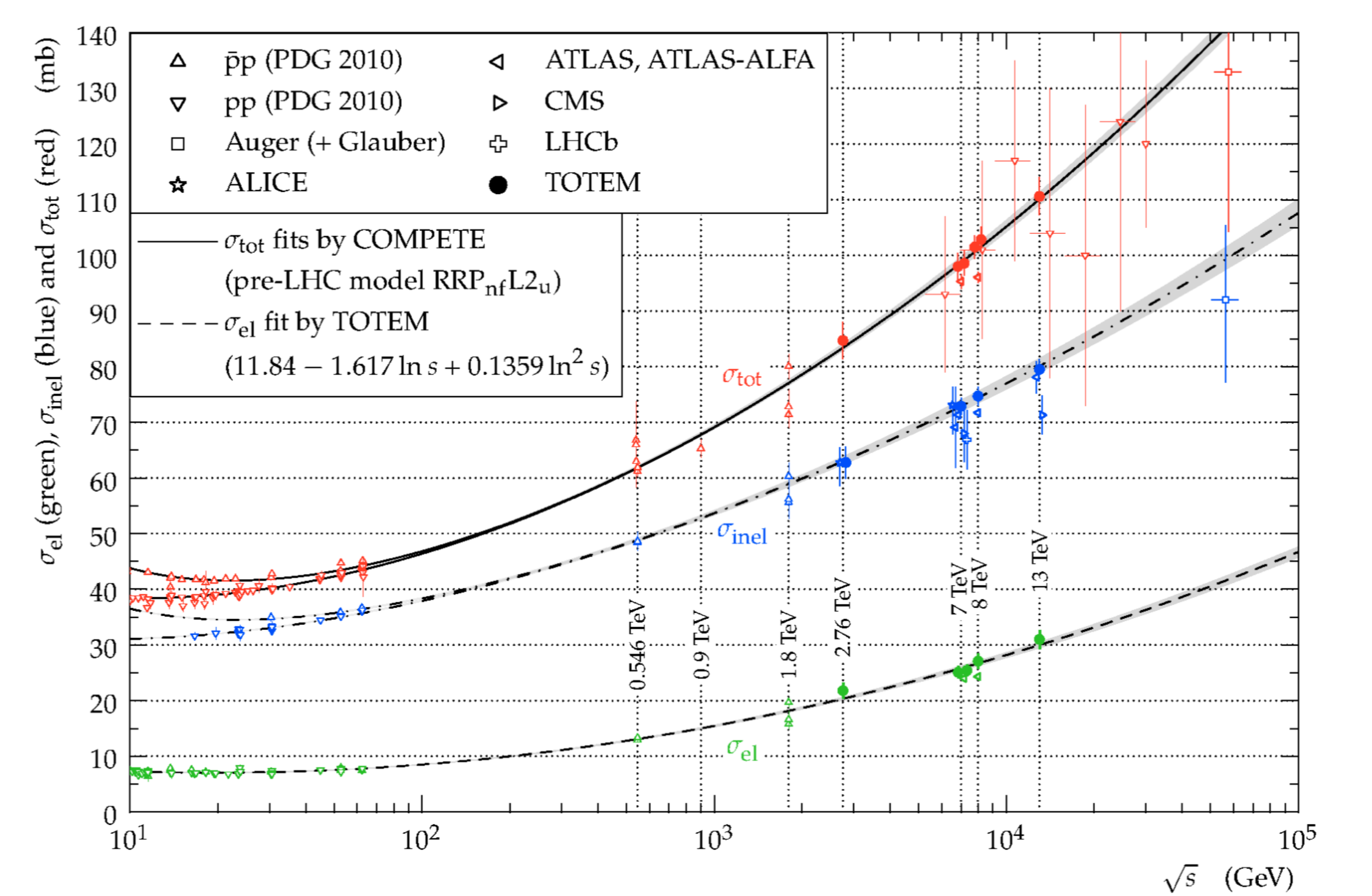}
\caption{The~energy dependence of the total, elastic and inelastic proton--proton 
cross sections.}
\label{fig1}
\end{figure}

Soon, the~perturbative Born approximation was used by Racah
\cite{rac} and preasymptotic terms with lower powers of $\ln \gamma $ were
calculated using this approach.
The main bulk of the total cross section is usually provided by hadronic
interactions. The~present experimental results from the energy behavior of the
cross sections of proton--proton interactions displayed in Figure~\ref{fig1}
\cite{tot, dr} demonstrate the approximately linear (or slightly stronger) 
increase with the logarithm of the energy.

The stronger regime, up to the square of the logarithm, is often used 
\cite{dr, royo} in practical fits. It is, in principle, admissible according 
to the famous Froissart bound \cite{froi} for purely hadronic interactions 
limited in space. Unfortunately, this theoretical bound is of no practical 
significance because it lies
much above experimental results due to a quite large numerical factor 
in front of the logarithm squared.

The large spatial extention of electromagnetic forces, in its turn, leads to~
the stronger energy increase of some inelastic processes. In view of such~ 
competition of electromagnetic and strong forces it is desirable to estimate  
at which energies and other experimental conditions these two contributions~ 
become of a comparable size and, also, to show where the ultraperipheral
processes start playing a role. 

\section{Simple Preasymptotic Estimates}
The high density of photons in electromagnetic fields surrounding charged 
colliding objects is responsible for strong increase of ultraperipheral
cross sections. 
The flux of photons is dominated by those carrying small fractions $x$ of the
nucleon energy. The~distribution of equivalent photons generated by a moving
nucleus with the charge $Ze$ (see, e.g., \cite{blp}) integrated over 
transverse momentum up to some value leads according to the method of
equivalent photons to the flux
\begin{equation}
\frac {dn}{dx}=\frac {2Z^2\alpha }{\pi x}\ln \frac {u(Z)}{x}.
\label{flux}
\end{equation}

The ultraperipherality parameter $u(Z)$ depends on the nature of colliding 
objects and differs numerically in various approaches
\cite{bgms, bb1, kn, bkn, kmr, geom, vyzh}. It has a physical meaning, which is the 
ratio of the maximum adoptable transverse momentum to the nucleon mass. 
It depends on charges $Ze$, energy, sizes (formfactors) and impact parameters
(the transverse distance between the centers) of colliding objects. The~impact 
parameters cannot be measured but, surely, should exceed the sum of the radii. This~requirement can be restated as a bound on the exchanged transverse momenta,
such that the objects are not destroyed but slightly deflected by the collision
so that no excitations or nuclear transitions happen.
The bound depends on their internal structure, i.e., on forces inside them.
These~forces are stronger for a proton than for heavy nuclei. Therefore protons
allow larger transverse momenta. The~quantitative estimates of the parameter
$u$ for different processes will be given below.

Besides the electron-positron pairs considered in Refs \cite{lali, rac}, other 
pairs of oppositely charged particles can be created in the two-photon collisions.
For example, pairs of muons produced in ultraperipheral collisions are 
observed at LHC \cite{at1, at2, at3}. The~light-by-light scattering described 
theoretically by the loop of charged particles is also detected at LHC
\cite{dent}. Some neutral bosons composed of quark-antiquark pairs can be
produced. This process is especially suitable for the compact theoretical
demonstration~\cite{geom} of $\ln ^3\gamma$-law (\ref{gamma}). 
The exclusive cross section of the production of the resonance $R$ 
in collisions of nuclei $A$ can be written as
\begin{equation}
\sigma _{AA}(R)=\int dx_1dx_2\frac {dn}{dx_1}\frac {dn}{dx_2}
\sigma _{\gamma \gamma }(R),
\label{e2}
\end{equation}
where the fluxes $dn/dx_i$ for the colliding objects 1 and 2
are given by Equation (\ref{flux}) and (see Ref. \cite{bgms})
\begin{equation}
\sigma _{\gamma \gamma }(R)=\frac {8\pi ^2\Gamma _{tot}(R)}{m _R}
Br(R\rightarrow \gamma \gamma )Br_d(R)\delta (x_1x_2s_{nn}-m_R^2).
\label{e3}
\end{equation}

Here $m_R$ is the mass of $R$, $\Gamma _{tot}(R)$ its total width and  
$Br_d(R)$ denotes the branching ratio to a considered channel of its decay.
$s_{nn}=(2m\gamma )^2, m$ is a nucleon mass.
The $\delta $-function approximation is used for resonances with small
widths compared to their masses.

The integrals in Equation (\ref{e2}) can be easily calculated so that one gets
the analytical formula
\begin{equation}
\sigma _{AA}(R)=\frac {128}{3}Z^4\alpha ^2Br(R\rightarrow \gamma \gamma )Br_d(R)
\frac {\Gamma _{tot}(R)}{m_R^3}\ln ^3\frac {2um\gamma }{m_R}.
\label{e4}
\end{equation}

The factor $2mu/m_R=1/\gamma _0$ defines the preasymptotic behavior of the 
ultraperipheral cross section of production of the resonance $R$.

It can be confronted with the formula for ultraperipheral production of muon
pairs in proton-- proton collisions derived in Equation (7) of \cite{vyzh}: 
\begin{equation}
\sigma (pp(\gamma \gamma )\rightarrow pp\mu ^+\mu ^-)=8\frac {28}{27}
\frac {\alpha ^4}{\pi m^2_{\mu }}\ln^3\frac {um\gamma }{m_{\mu }}.
\label{ppvz}
\end{equation}

The energy dependence of both processes is the same for $m_R=2m_{\mu }$ 
as expected. The~preasymptotic behavior is determined by the factor $um/m_R$.
The asymptotic limit is reached at 
\begin{equation}
\gamma \gg m_R/2um,
\label{asym}                 
\end{equation}
where the terms increasing slower than $\ln ^3\gamma $ can be neglected.
\
The parameter $u$ is the least precisely determined element of the whole 
approach. The~careful treatment of formfactors of protons and nuclei with 
account of the photon virtuality (see also Refs \cite{bb1, bkn} where the
problem was treated in more detail) and the suppression factors \cite{vyzh} lead
 to its values $u_{pp}\approx 0.2$ for $pp$ and $u_{PbPb}\approx 0.02$ for
$PbPb$-collisions within the factors about 1.5 which depend on the particular
shape of the formfactors (see \cite{vyzh}). In what follows, motivated by these 
results I use the parameters $u$ obtained in Ref. \cite{vyzh}. They favor 
ultraperipheral processes at lower energies than the extremely strong ad-hoc 
requirements of the cutoff of the impact parameters imposed in \cite{kmr} and 
used in \cite{geom} which give rise to about 4 times smaller values of $u$, i.e.,
 to the higher lying (on the energy scale) asymptotics. 

Taking into account these caveats, one can confront the values and energy 
dependences of experimentally measured cross sections of inelastic 
$pp$-interactions and the relative contribution of ultraperipheral processes 
in them. The~values of the inelastic cross section shown in Figure \ref{gamma} are rather 
well approximated in the energy interval from 60 GeV to 13 TeV
by the expression
\begin{equation}                  
\sigma _{inel}(s)=8.2\ln (1.37\sqrt s) \; mb,
\label{inel}
\end{equation}
where $\sqrt s$ is in GeV. 

Let us consider the channel with a single $\pi^0$ produced among all inelastic 
channels and compare it with the expression for the ultraperipheral cross 
section for $\pi ^0$-production. The~multiplicity distribution is well 
described in the considered energy interval by a composition of the negative 
binomial distributions (NBD) \cite{dnec} with the average multiplicity $\bar n$ and 
the dispersion determined by $k$. It~is dominated by a single NBD for events 
with low multiplicities. The~probability to get the inelastic process
with a charged pion produced is equal to
\begin{equation}
P(\pi ^{\pm })=\bar n\left (1+\frac {\bar n}{k}\right )^{-k-1}.
\label{pi}                                                          
\end{equation}

It is twice smaller for a neutral pion, such that 
$P(\pi ^{0})\approx$ 4 $\times$ $10^{-3}$ for $\bar n$ = 13, $k$ = 4.4 
at the intermediate 
(for the chosen interval) energy 
1.8 TeV (see \cite{dnec}). The~product $P(\pi ^{0})\sigma _{inel}(s)$ must be 
compared with Equation~(\ref{e4})\footnote{The $\pi ^0$-production in 
2$\gamma $-collisions was originally suggested by Low \cite{low}.}
 for $Z=1$. The~preasymptotic factors in logarithms
are very close to one another (1.37~in~(\ref{inel}) compared to 1.48 in 
(\ref{e4}) for $u$ chosen according to Ref.~\cite{vyzh}).
Thus, the~cubic equation obtained from the equality of the product 
$P(\pi ^{0})\sigma _{inel}(s)$ to Equation (\ref{e4})
reduces approximately to the quadratic one. The~energy $s_0$, at which the
cross section of the ultraperipheral production of a single $\pi ^0$ becomes
equal to its partial cross section due to hadronic interactions,
can be estimated from the~equality
\begin{equation}
2.7\times 10^{-9}\ln ^2 \sqrt {s_0}\approx 3.28\times 10^{-2}.
\label{uo}                                                          
\end{equation}

{The photon fluxes for $pp$ collisions with $Z=1$  are not strong (see Equations
(\ref{flux}) and (\ref{e2})). Therefore, the~factor in front of 
the ultraperipheral contribution on the left hand side is extremely small. 
One concludes that these expressions can become equal only at the unrealistically 
high energy \mbox{$\sqrt {s_0}\approx e^{3500}$}~GeV.} At first sight, it seems hopeless 
to measure such processes at the present energies in $pp$-collisions.
To enlarge their share, one should try to impose some special experimental 
cutoffs. Fortunately, there are distinctive features which can help~
in choices of such events. In particular, the~ultraperipherally created ~
neutral pions move slowly, decay to two photons with energies 67.5 MeV 
and are strongly concentrated near central rapidity. The~whole process looks
like the light-by-light scattering at the specific $\pi ^0$ energy.~
Surely, the~fiducial cross sections of both ultraperipheral and hadronic 
interactions would be strongly diminished.

The optimism is supported by studies \cite{vyzh} of ultraperipheral production 
of $\mu ^+\mu ^-$ pairs. The~ultraperipheral cross section (\ref{ppvz}) at 
13 TeV is equal\footnote{Please note that it includes the $\ln ^3\gamma$-factor 
which is about 700.} to 0.22 $\mu $b. It is much smaller than
the inelastic cross section of 80 mb. Further cuts on the invariant mass
of the $\mu ^+\mu ^-$ pair, on the muon transverse momentum and  
pseudorapidity reduce its value to 3.35 pb. If corrected for absorptive effects~\cite{dysh} it gives \mbox{3.06 $\pm$ 0.05 pb}. The~chosen cuts coincide with
those imposed in studies of the ATLAS collaboration~\cite{at1} which
lead to the value 3.12 $\pm $ 0.07 (stat.) $\pm $ 0.10 (syst.) pb. The~Monte Carlo
program~\cite{hlkr} which incorporates both ordinary and ultraperipheral 
processes predicts 3.45 $\pm $ 0.06 pb. Theoretical results are in agreement
with experimental data and show that ultraperipheral processes dominate
over other sources in this fiducial volume. Analogous conclusions
were obtained for lead-lead collisions \cite{vyzh}. The~measured fiducial cross
sections are on the $\mu $b scale compared to pb's for~$pp $.

The creation of a $\pi ^0$ in collisions of heavy nuclei is strongly enhanced
by the factors $Z^4=4.5\times 10^7$ for $PbPb$ or $3.9\times 10^7$ for $AuAu$
collisions which must appear on the left hand side of the equation analogous to
(\ref{uo}). That makes it of 
 comparable size to the hadronic contribution on the right hand side even if 
the larger nuclear cross sections (of the order of the geometrical size
about 1500 mb) are inserted there. The~factor of the stronger
energy increase of ultraperipheral processes becomes decisive now.
Taking into account the value of $u\approx 0.02$ applicable to heavy nuclei, 
the effect could become observable even at comparatively low energies of 
NICA (with $\gamma =4.5 - 6$) because the preasymptotic threshold 
(\ref{asym})\footnote{The previous estimate of the preasymptotic threshold 
\cite{geom} was 4 times larger as mentioned above and excluded NICA~energies.}
asks for $\gamma >3.6$. Again, photons with energies
67.5 MeV in the central rapidity region can be looked for as a signature
for decays of slowly moving neutral pions produced in ultraperipheral
collisions. The~threshold for heavier resonances is proportional to their
masses (see (\ref{asym})) and, therefore, moves to higher energies.
Quantitative comparison can be done after the Monte Carlo program 
for the exclusive resonance production similar to the STARlight program 
for $\mu ^+\mu ^-$ processes \cite{knsgb} is elaborated and helps in search
of the proper fiducial phase space volume.

Special attention should be paid at NICA to the production of 
$e^+e^-$-pairs\footnote{Let us mention that at the LHC this process was studied 
in $PbPb$ collisions at $\sqrt s$ = 2.76 TeV by the ALICE collaboration 
\cite{alice}.}. If observed, it would show that ultraperipheral processes
survive even at comparatively low NICA energies. Their cross section can be 
estimated according to the Racah formula \cite{rac, bkn} which accounts for the
terms that increase slower with energy in the perturbative Born approximation:
\begin{equation}
\sigma =\frac {28}{27\pi }\frac {Z^4\alpha ^4}{m_e^2}[L^3-2.2L^2+3.84L-1.64],
\label{racah}
\end{equation}
where $m_e$ is the electron mass, $L=\ln \gamma ^2$. For the energy 
4.5 GeV per nucleon in $AuAu$ collisions one gets very large cross section 
$\sigma \approx $ 1~kb exceeding the  values for geometrical
estimates of hadronic processes at low impact parameters. The~small 
electron mass plays a crucial role. The~$L^2$ term is negative and 
about 30$\%$ of the main $L^3$-term which dominates asymptotically.
Even at the LHC energy 5 TeV it is about 10$\%$. Moreover, the~$L^2$-term of 
the Coulomb
corrections to the Racah formula is also negative \cite{iss, lms, geku}
\begin{equation}
\sigma _C =-\frac {56}{9\pi }\frac {Z^4\alpha ^4}{m_e^2}f(Z)L^2,
\label{coul}
\end{equation}
where
\begin{equation}
f(Z)=(Z\alpha )^2\sum _{n=1}^{\infty }\frac {1}{n(n^2+(Z\alpha )^2)}.
\label{fZ}
\end{equation}

These corrections are large at NICA energies. Effectively, they
can be accounted by the replacement of the coefficient $-$2.2 in front of the 
$L^2$ term by $-$4.2. The~preasymptotic ``threshold'' determined by the requirement
that $L^3$-contribution is larger than that of the whole $L^2$-term is shifted 
to higher energies and poses a problem at NICA. 

However, for quantitative treatment, the~proper account of the nuclei structure 
should be incorporated phenomenologically in these perturbative calculations. 
The nuclei were treated in Refs~\cite{lali, rac} as pointlike objects.
The strong dependence of preasymptopia on their structure (the parameter $u$)
and on masses of created particles is claimed according to the method of 
equivalent photons. If Equation (\ref{e4}) is used for estimates of the 
$L^2$-term with $m_R=2m_e$ one finds quite large positive coefficients of 
values about 10 in front of it. That would compensate the negative 
contributions discussed above and shift the ``threshold'' to lower energies.
In any case, the~estimated
values of the cross section are still quite optimistic. Equation~(\ref{racah}) 
cannot be trusted at $\gamma $ approaching 3 where the
second term becomes equal to the first one. The~background 
from collisions at low impact parameters at NICA must be much smaller than at 
LHC energies. Surely, special detectors must be installed to separate 
electrons from other products of the collisions.

The muon pairs cannot be created ultraperipherally at NICA energies.
The corresponding preasymptotic factor $um/m_{\mu}$ in Equation~(\ref{ppvz})
becomes about 200 times smaller than $um/m_e$ for the $e^+e^-$-case. 
In other words, it means that the parameter $\gamma _0$ becomes 200 times
larger. That leads to the negative $L^2$-term, i.e., to a higher ``threshold''.

Let me mention at the very end that the planned NICA energies are very close 
to those at the FAIR facilities. Thus, all conclusions concerning the 
ultraperipheral processes are equivalent in both cases.

\section{Conclusions}
Electromagnetic fields of colliding charged particles are the reason for the 
fast increase of their ultraperipheral cross sections with energy.
These cross sections are strongly enlarged by the nuclear charge $Z$ for
interactions of heavy ions. The~exclusive production of resonances in these
processes is compared with cross sections of ordinary hadronic 
interactions for $pp$ and $PbPb$ high energy collisions. It is small in $pp$ 
and quite noticeable in $PbPb$ processes due to the $Z^4$-factor. Using the
parameters of preasymptotic estimates borrowed from \cite{vyzh}, it is argued 
that the ultraperipheral processes $AA\rightarrow AA\pi ^0$ and 
$AA\rightarrow AAe^+e^-$ can be observed at pernucleon energies higher than 
3.6~GeV.

\vspace{6pt}

{\bf Acknowledgment}

This work was supported by the RFBR project 18-02-40131 and RAN-CERN program.

I am grateful to C. Bertulani for references \cite{bb1, bkn}.


\begin{thebibliography}{999}

\bibitem{lali}
Landau, L.D.; Lifshitz, E.M.
On the production of electrons and positrons by a collision of two particles.
{\em Physikalische Zeitschrift der Sowjetunion} {\bf 1934},~{\em 6},~244.

\bibitem{fermi}
Fermi, E.
{On the Theory of the impact between atoms and electrically charged particles}.
{\em Zeitschrift Für Physik} {\bf 1924},~{\em 29},~315--327.

\bibitem{weiz}
Weizs\"{a}cker, C.F.V.
{Radiation emitted in collisions of very fast electrons}.
{\em Zeitschrift Für Physik} {\bf 1934},~{\em 88},~612--625.



\bibitem{will}
{Williams, E.J. Nature of the high energy particles of penetrating radiation and status of ionization and radiation formulae.
\emph{Phys. Rev.} \textbf{1934},~\emph{45},~729--730.}


\bibitem{rac}
Racah, G. Sulla nascita di coppie per urti di particelle elettrizzate. 
{\em Nuovo Cim.} {\bf 1937},~{\em 14},~93.

\bibitem{tot}
Aspell, P. et al. [The TOTEM Collaboration] First measurement of elastic, inelastic and total cross-section at $\sqrt {s}= 13$ TeV by TOTEM and overview of cross-section data at LHC energies.
{\em arXiv} \textbf{2017}, arXiv:1712.06153.


\bibitem{dr}
Dremin, I.M. Some Recent Results on High-Energy Proton Interactions.
{\em Particles} {\bf 2019},~{\em 2},~57--69.

\bibitem{royo}
Royon, C. Total cross section and particle production in soft and hard processes at the LHC.
{\em arXiv} \textbf{2019}, arXiv:1909.12696.


\bibitem{froi}
Froissart, M. Asymptotic behavior and subtractions in the Mandelstam representation.
{\em Phys. Rev.} {\bf 1961},~{\em 123},~1053--1057.

\bibitem{blp}
Berestetsky, V.B.; Lifshitz, E.M.; Pitaevsky, L.P.
\emph{Kvantovaya Electrodinamika};
Fizmatlit: Moscow, Russia, 2001.

\bibitem{bgms}
Budnev, V.M.; Ginzburg, I.F.; Meledin, G.V.; Serbo, V.G. The two-photon particle production mechanism. Physical problems. Applications. Equivalent photon approximation. 
{\em  {Phys. Rep. C}} {\bf 1975},~{\em 15},~181--282. 


\bibitem{bb1}
Baur, G.; Bertulani, C.A. $\gamma$--$\gamma$ physics with peripheral relativistic heavy ion collisions.
{\em Z. Phys. A} {\bf 1988},~{\em 330},~77--81.



\bibitem{kn}
Klein, S.R.; Nystrand, J. Interference in exclusive vector meson production in heavy-ion collisions.
{\em Phys. Rev. Lett.} {\bf 2000},~{\em 84},~2330--2333.


\bibitem{bkn}
Bertulani, C.A.; Klein, S.R.; Nystrand, J. Physics of ultra-peripheral nuclear collisions.
{\em Ann. Rev. Nucl. Part. Sci.} {\bf 2005},~{\em 55},~271--310.


\bibitem{kmr}
Khoze, V.A.; Martin, A.D.; Ryskin, M.G. Exclusive vector meson production in heavy ion collisions.
{\em arXiv} \textbf{2019}, {arXiv:1902.08136}.



\bibitem{geom}
Dremin, I.M. Geometry of ultraperipheral nuclear collisions. {\em  Int. J. Mod. Phys. A} {\bf 2019},~{\em 34},~1950068.



\bibitem{vyzh}
Vysotsky, M.I.; Zhemchugov, E.V. Equivalent photons in proton–proton and ion–ion collisions at the Large Hadron Collider.
\emph{Phys. Usp.} {\bf 2019},~{\em 189},~975--984.



\bibitem{at1}
Aaboud, M. et al. [ATLAS Collaboration] Measurement of the exclusive gamma gamma-> mu (+) mu (-) process in proton-proton collisions at root s= 13 TeV with the ATLAS detector.
{\em  Phys. Lett. B} {\bf 2018},~{\em 777},~303--323. 




\bibitem{at2}
Aaboud, M. et al. [ATLAS collaboration] Expected performance of the ATLAS inner tracker at the High-Luminosity LHC.
ATLAS-CONF-2016-025. 2016.


\bibitem{at3}
Arratia, M. Ultra-peripheral Collisions with the ATLAS Detector.
{\em arXiv} {\bf 2016}, arXiv:1611.05145.


\bibitem{dent}
d'Enterria, D. et al. [CMS Collaboration] Evidence for light-by-light scattering in ultraperipheral PbPb collisions at sNN= 5.02 TeV.
{\em Nucl. Phys. A} {\bf 2019},~{\em 982},~791--794.


\bibitem{dnec}
Dremin, I.M.; Nechitailo, V.A. Soft multiple parton interactions as seen in multiplicity distributions at Tevatron and LHC.
{\em Phys. Rev. D} {\bf 2011},~{\em 84},~034026.



\bibitem{low}
Low, F.E. Proposal for Measuring the $\pi ^0$ Lifetime by $\pi ^0$ Production in Electron-Electron or Electron-Positron Collisions.
{\em Phys. Rev.} {\bf 1960},~{\em 120},~582--583. 




\bibitem{dysh}
Dyndal, M.; Schoeffel, L. The role of finite-size effects on the spectrum of equivalent photons in proton-proton collisions at the LHC.
{\em Phys. Lett. B} \textbf{2015}, {\em 741}, 66--70.



\bibitem{hlkr}
Harland-Lang, L.A.; Khoze, V.A.; Ryskin, M.G. Exclusive physics at the LHC with SuperChic 2.
{\em Eur. Phys. J. C} {\bf 2016},~{\em 76},~9. 



\bibitem{knsgb}
Klein, S.R.; Nystrand, J.; Seger, J.; Gorbunov, Y.; Butterworth, J. STARlight: A Monte Carlo simulation program for ultra-peripheral collisions of relativistic ions.
{\em Comput. Phys. Commun.} {\bf 2017},~{\em 212},~258--268.


\bibitem{alice}
Abbas, E. et al. [The ALICE Collaboration] Charmonium and e + e − pair photoproduction at mid-rapidity in ultra-peripheral Pb–Pb collisions at sNN−−−√=2.76 TeV.
{\em Eur. Phys. J. C} {\bf 2013},~{\em 73},~2617.



\bibitem{iss}
Ivanov, D.Y.; Schiller, A.; Serbo, V.G. Large Coulomb corrections to the e+ e− pair production at relativistic heavy ion colliders.
{\em  Phys. Lett. B} {\bf 1999},~{\em 454},~155--160.



\bibitem{lms}
Lee, R.N.; Milstein, A.I.; Serbo, V.G. Structure of the Coulomb and unitarity corrections to the cross section of e+ e− pair production in ultrarelativistic nuclear collisions.
{\em Phys. Rev. A} {\bf 2002},~{\em 65},~022102.


\bibitem{geku}
Gevorkyan, S.R.; Kuraev, E.A. Lepton pair production in relativistic ion collisions to all orders in Zα with logarithmic accuracy
{\em J. Phys. G} {\bf 2003},~{\em 29},~1227.



\end{thebibliography}
\end{document}